\documentclass[pra,twocolumn,showpacs,preprintnumbers,amsmath,amssymb,superscriptaddress]{revtex4-1}

\usepackage{graphicx}
\usepackage{dcolumn}
\usepackage{bm}
\usepackage{epsfig}
\usepackage{amsmath}
\usepackage{amssymb}
\usepackage{color}
\usepackage{bbm}

\usepackage[colorlinks=true,citecolor=blue,linkcolor=blue,urlcolor=blue]{hyperref}

\begin{document}

\title{Ultrastrong coupling phenomena beyond the Dicke model}

\author{Tuomas~Jaako}
\affiliation{Vienna Center for Quantum Science and Technology,
Atominstitut, TU Wien, Stadionallee 2, 1020 Vienna, Austria}
\author{Ze-Liang~Xiang}
\affiliation{Vienna Center for Quantum Science and Technology,
Atominstitut, TU Wien, Stadionallee 2, 1020 Vienna, Austria}
\author{Juan Jos\'e Garcia-Ripoll}
\affiliation{Instituto de F\' isica Fundamental, IFF-CSIC, Calle Serrano 113b, Madrid E-28006, Spain}
\author{Peter~Rabl}
\affiliation{Vienna Center for Quantum Science and Technology,
Atominstitut, TU Wien, Stadionallee 2, 1020 Vienna, Austria}

\date{\today}

\begin{abstract}
We study effective light-matter interactions in a circuit QED system consisting of a single $LC$ resonator, which is coupled symmetrically to multiple superconducting qubits. Starting from a minimal circuit model, we demonstrate that in addition to the usual collective qubit-photon coupling the resulting Hamiltonian contains direct qubit-qubit interactions, which have a drastic effect on the ground and excited state properties of such circuits in the ultrastrong coupling regime. In contrast to a superradiant phase transition expected from the standard Dicke model, we find an opposite mechanism, which at very strong interactions completely decouples the photon mode and projects the qubits into a highly entangled ground state. These findings resolve previous controversies over the existence of superradiant phases in circuit QED, but they more generally show that the physics of two- or multi-atom cavity QED settings can differ significantly from what is commonly assumed.
\end{abstract}

\pacs{ 42.50.Pq, 	
05.30.Rt, 	
85.25.-j 
        }
\maketitle

%
%


\section{Introduction}
The Dicke model (DM) is frequently used in atomic and solid-state systems as a minimal model to describe phenomena related to the collective coupling of many emitters to a single radiation mode~\cite{Dicke,BrandesPR2005}.  
When extended to the ultrastrong coupling regime, where the collective coupling becomes comparable to the atomic and optical frequencies, the ground state of the DM undergoes a phase transition into a superradiant state \cite{HeppLieb, WangHioe, Emary}. This state is characterized by a non-vanishing field expectation value and a uniform atomic polarization and is commonly considered the hallmark of ultrastrong coupling physics. While this superradiant phase transition (SRT) is well understood and has been observed with engineered Hamiltonians in driven atomic systems~\cite{Esslinger,Baden2014,KlinderPNAS2015}, the validity of the Dicke model for describing also the ground states of equilibrium cavity QED systems is still subject of ongoing debates~\cite{Rzazewski, Knight,Keeling2007,DeLiberato,VukicsPRL2014,Griesser2016}. 
This question has regained considerable interest with the development of circuit QED systems~\cite{BlaisPRA2004,WallraffNature2004,DevoretAnnPhys2007,YouNature2011}, where superconducting two-level systems are strongly coupled to microwave photons. In particular, it has been argued~\cite{CiutiNatComm2010} that the equivalent of the so-called $A^2$ term---which supposedly prevents the SRT in atomic systems~\cite{Rzazewski}---does not play a crucial role in these artificial circuit QED devices. However, these predictions have also been questioned on very general grounds~\cite{ViehmannPRL2009}, or based on concrete models~\cite{Leib2012}. While recently the ultrastrong coupling regime for single qubits has been experimentally achieved \cite{NiemczykNatPhys2010,FornDiazPRL2010,Chen2016,Yoshihara2016,FornDiaz2016}, the nature of the ground states of collective circuit QED systems still remains open. 

In this work we investigate a multiqubit generalization of a circuit QED system, where $N$ charge qubits are coupled symmetrically to a single microwave mode. In the limit of weak coupling the system reduces to the standard DM and thus such circuits have been proposed for studying the superradiant phase transition~\cite{CiutiNatComm2010,ChenPRA2007,LambertPRB2009}, which is expected to occur when either the coupling strength or the number of qubits is increased. The important finding of this work is that the full Hamiltonian for this circuit necessarily contains additional direct qubit-qubit interactions, which have been ignored in many previous studies, but become non-negligible as one approaches the ultrastrong coupling regime. The analysis of the ground state properties of this extended Dicke model (EDM) reveals surprising new effects, which completely contradict our existing intuition about light-matter interaction. Instead of undergoing a transition into a superradiant phase with increasing coupling strength, the system first gradually evolves into a hybridized qubit-photon state, but without broken symmetry. At even higher interaction strengths, an opposite effect can take place, where the photonic component of the ground state completely decouples, while the qubits collapse into a highly entangled Dicke state with vanishing dipole moment. We explain these findings with an effective low-energy theory, which reveals the existence of separate manifolds with an exponentially large number of nearly degenerate states. Together with the high degree of entanglement, this feature makes the EDM considerably more involved than the well-studied DM and imposes many new challenges for theoretical and experimental research on ultrastrong coupling physics.

The analysis presented in this work is primarily focused on a minimal model for a charge-coupled circuit, which is motivated by previous studies on this subject and allows a simple physical interpretation of the predicted effects. It also establishes a direct connection to  
the single-mode Hopfield model that is used to describe ultrastrong coupling effects in other electrically coupled cavity QED settings~\cite{Hopfield1958,TodorovPRL2010,TodorovPRX2014}, or related models for coupled quantum dots~\cite{CottetPRB2015}, and shows the broader scope of our findings. However, the EDM can also be derived for inductively coupled circuits (although not for all~\cite{Bamba2016}) based on the flux qubit design. For this type of a qubit, the ultrastrong coupling regime is already experimentally accessible today~\cite{Yoshihara2016,FornDiaz2016} and using the tuneability of the coupling \cite{Yoshihara2016,FornDiaz2016,PeropadrePRL2013} in these setups, it should be possible to dynamically test the predicted new effects for $N=2$ or more qubits, over a wide range of coupling strengths, including the scaling of the qubit-qubit interaction strength, light-matter decoupling and ground state properties. 


The remainder of the paper is structured as follows. In Sec.~\ref{sec:Model} we derive the exact Hamiltonian for a minimal collective circuit QED systems and discuss in Sec.~\ref{sec:USC} the meaning of superradiant states, first for the example of a single qubit. In Sec.~\ref{sec:IDM} we then analyze the ground state properties of the collective model and explain the origin of the light-matter-decoupling effect. In Sec.~\ref{sec:FQ} we briefly describe alternative circuit implementations based on the flux qubit design and demonstrate the robustness of the observed effects with respect to experimental imperfections. Finally, Sec.~\ref{sec:Conclusions} summarizes the main conclusions of this work.

\begin{figure}[t]
  \centering
    \includegraphics[width=0.48\textwidth]{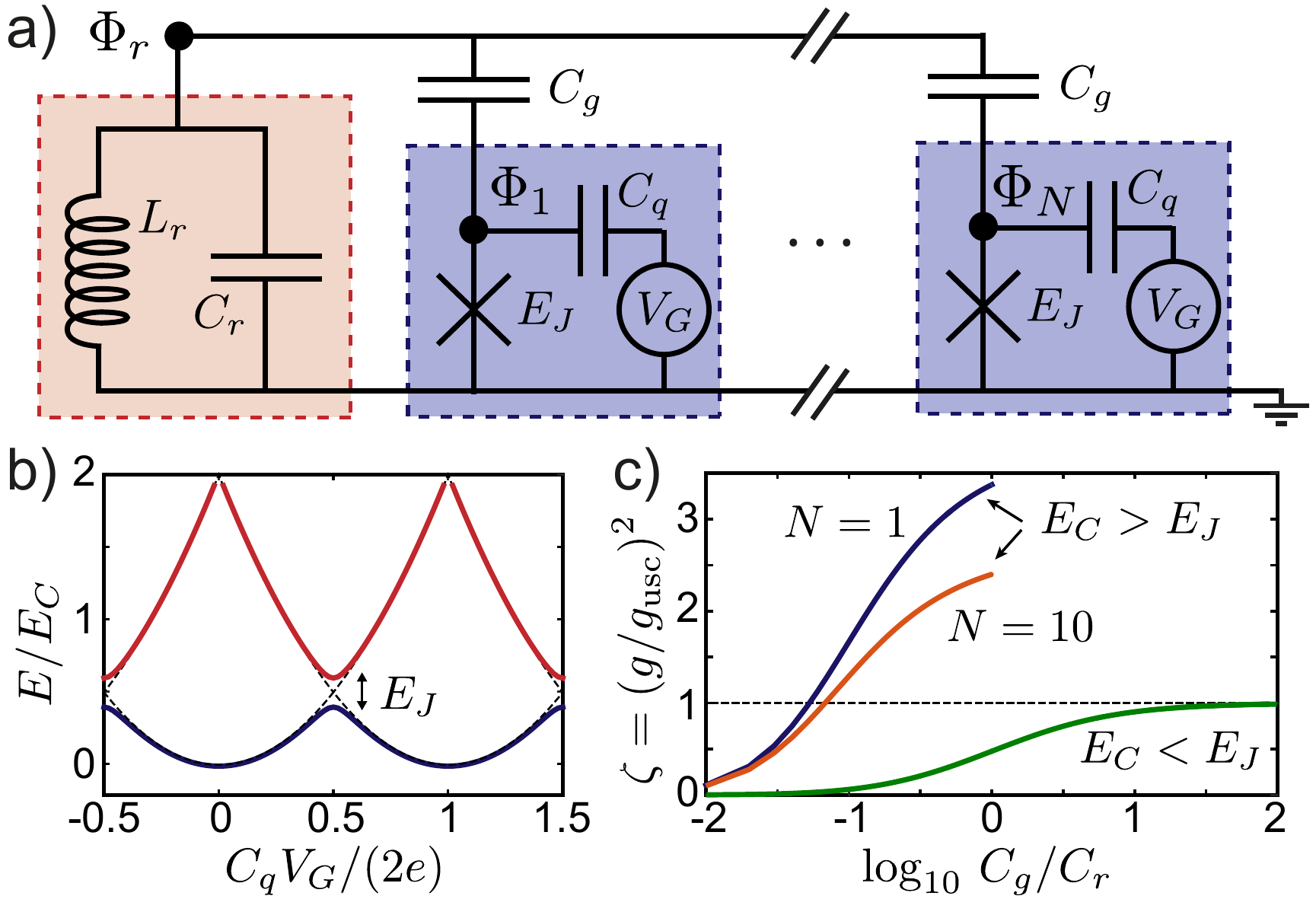}
      \caption{a) Circuit model for $N$ charge qubits coupled to a single lumped-element resonator. b) Sketch of the energy levels of a single charge qubit as a function of the applied gate voltage and for $E_C\gg E_J$. c) Plot of the coupling parameter $\zeta=(g/g_{\rm usc})^2$ as a function of $C_g/C_r$ and a specific set of circuit parameters $C_r = 10\,\mathrm{fF} $, $ C_q = 0.5\,\mathrm{fF} $, and $L_r = 1\,\mathrm{nH}$. For the two upper lines $E_J/h = 2\,\rm{GHz}$ and the condition $E_J/E_C<1$ is fulfilled over the whole range of plotted values. The lowest line represents the transmon limit ($E_J/E_C>1$) given in Eq.~\eqref{eq:CouplingLimit1} for $N=1$.}
      \label{Fig1Setup}
\end{figure}

\section{Collective circuit QED}\label{sec:Model}

We consider a superconducting circuit as depicted in Fig.~\ref{Fig1Setup} a), where $N$ charge qubits~\cite{MakhlinRMP2001,YouNature2011} are coupled symmetrically to a single mode lumped element resonator with capacitance $C_r$ and inductance $L_r$. 
 Each qubit is represented by a Cooper pair box with capacitance $C_q$ and Josephson energy $E_J$ and coupled to the $LC$ resonator via an additional capacitance $C_g$. The whole circuit is described by the Lagrangian 
\begin{equation}\label{eq:Lagrangian}
\mathcal{L}=\frac{C_r\dot \Phi_r^2}{2} - \frac{\Phi_r^2}{2L_r} +\sum_{i=1}^N  \left[ \mathcal{L}_q(\Phi_i,\dot \Phi_i) +\frac{C_g}{2} \left( \dot \Phi_r-\dot \Phi_i\right)^2\right],
\end{equation}
where $\Phi_\eta(t)= \int_{-\infty}^t  V_\eta(s) ds$ is the generalized flux~\cite{DevoretLesHouches} associated with the voltage $V_\eta$ at each node, $\eta=r,1,\dots,N$ [see Fig.~\ref{Fig1Setup} a)].  The Lagrangian for each qubit is 
\begin{equation}
\mathcal{L}_q(\Phi,\dot \Phi)= \frac{C_q}{2}( \dot \Phi-V_G)^2 + E_J  \cos\left(\frac{\Phi}{\Phi_0}\right),
\end{equation}
where $V_G$ is the applied gate voltage and  $\Phi_0=\hbar /(2e)$ is the reduced flux quantum. We emphasize that apart from the coupling of each individual qubit to the cavity, our model in Eq.~\eqref{eq:Lagrangian} does not contain any direct capacitive interactions among the qubits themselves.

To obtain a quantized circuit model, we follow the standard quantization procedure and replace the $\Phi_\eta$ and the conjugate node charges $Q_\eta =\partial \mathcal{L}/\partial \dot \Phi_\eta$ by operators obeying $[\Phi_\eta,Q_{\eta'}]=i\hbar \delta_{\eta\eta'}$. For the resulting Hamilton operator $H=\sum_\eta Q_\eta V_\eta - \mathcal{L}$ 
we then obtain (see App.~\ref{App:CircuitHamiltonian})
\begin{equation}\label{eq:FullHamiltonian}
\begin{split}
H=& \frac{Q_r^2}{2\bar C_r} + \frac{\Phi_r^2}{2L_r} +\sum_{i=1}^N \left [H_q^i 
+\frac{Q_r \mathcal{Q}_i}{\bar C_g}\right] +\sum_{i\neq j} \frac{\mathcal{Q}_i \mathcal{Q}_j}{2\bar C_{qq}},
\end{split}
\end{equation}
where  $\mathcal{Q}_i=Q_i+C_qV_G$ is the displaced charge and 
\begin{equation}\label{eq:Hqubit}
H_q^i= \frac{\mathcal{Q}_i^2}{2\bar C_q} -E_J   \cos\left(\frac{\Phi}{\Phi_0}\right)
\end{equation}
is the Hamiltonian for an individual qubit. The modified capacitances that appear in Eqs.~\eqref{eq:FullHamiltonian} and~\eqref{eq:Hqubit} are given by $ \bar C_q=\bar C^2/[C_r+C_g+(N-1)C_gC_q/(C_q+C_g)]$, $\bar C_r = \bar C^2/(C_q+C_g)$, $\bar C_g=\bar C^2/C_g $, $\bar C_{qq}=(C_g+C_q)\bar C^2/C_g^2$ and $\bar C^2 = C_gC_r +C_q(C_r+N C_g)$.  

Equation~\eqref{eq:FullHamiltonian} first of all shows that the coupling of individual qubits to a common resonator mode renormalizes the bare resonator and qubit energies. This is analogous to the effect of the $A^2$-term, which in atomic cavity QED systems increases the photonic field energy proportional to the number of atoms and can thereby prevent a SRT~\cite{Rzazewski}. Interestingly, here the coupling \emph{lowers} the charging energies, i.e., $\bar C_{r,q}> C_{r,q}$, which is exactly the opposite effect and would at first sight favor a SRT.  
However, from the last term in Eq.~\eqref{eq:FullHamiltonian} we also see that the Legendre transformation from voltage to charge variables introduces additional direct qubit-qubit interactions, the effect of which we will analyze in the following.

In a final step we express $Q_r= Q_0^r(a^\dag +a)$ and $\Phi_r= i Q_0^r/(\bar C_r \bar \omega_r) (a^\dag -a)$ in terms of annihilation and creation operators, $a$ and $a^\dag$, where $\bar \omega_r=1/\sqrt{L_r\bar C_r}$ and $Q_0^r=\sqrt{\hbar \bar C_r \bar \omega_r/2}$.  We further restrict each qubit to the two lowest states $|0\rangle$ and $|1\rangle$ (at this stage without further justification), which are separated by an energy $\hbar\bar \omega_q$. For the examples considered below we can then  write $H_q^i\simeq \hbar \bar \omega_q \sigma_z^i/2$ and $\mathcal{Q}_i\simeq Q_0^q \sigma_x^i$, where $Q_0^q=\langle 1|\mathcal{Q}|0\rangle$ and the $\sigma_k$ are the usual  Pauli operators.  Under these assumptions and by introducing collective spin operators $S_k = \frac{1}{2}\sum_i \sigma_k^i$, we finally obtain $(\hbar=1)$
\begin{equation}\label{eq:TLAHamiltonian}
H\simeq \bar \omega_r a^\dag a + \bar \omega_q  S_z  + g(a + a^\dag)S_x   + D S_x^2, 
\end{equation}
where $g=2Q_0^rQ_0^q/(\hbar \bar C_g)$ and $D=2(Q_0^q)^2/(\hbar \bar C_{qq})$. Eq.~\eqref{eq:TLAHamiltonian} thus represents the minimal model for $N\geq 2$ qubits that are coupled capacitively to a single microwave mode. Below we will show that identical models can be derived for basic flux-coupled circuits, these models are also discussed in other cavity QED implementations~\cite{TodorovPRX2014,TodorovPRL2010}. In the absence of the last term, Eq.~\eqref{eq:TLAHamiltonian}  reduces to the standard DM, which exhibits a superradiant ground state for couplings $\sqrt{N}g\geq  g_{\rm usc}$~\cite{Emary}, where we use $g_{\rm usc}= \sqrt{\bar\omega_r \bar\omega_q}$ to define the onset of the ultrastrong coupling regime for a single qubit.   However, since they have the same physical origin the qubit-qubit interaction strength $D$ and the qubit-resonator coupling $g$ are not independent. In view of $\bar C_{qq}= \bar C_g^2/\bar C_r$ in Eq.~\eqref{eq:FullHamiltonian}, we obtain the exact relation
 \begin{equation}\label{eq:D}
D= \frac{g^2}{\bar \omega_r},
\end{equation}
which demonstrates the significance of this additional term in the limit of very strong interactions.

\section{Ultrastrong coupling}\label{sec:USC}

Before we proceed with the multi-qubit model let us first evaluate the value of the single-qubit coupling parameter $\zeta=(g/g_{\rm usc})^2$ that can actually be achieved for a specific circuit design. We first do so for the frequently used transmon qubit~\cite{KochPRA2007}, which is operated in the regime $E_J\gg E_C=e^2/(2\bar C_q)$ and $V_G=0$. In this case the two lowest eigenstates are well approximated by harmonic oscillator states with $\bar \omega_q\simeq \sqrt{8 E_CE_J}$ and $Q_0^q\simeq\sqrt{\hbar \bar C_q \bar \omega_q/2}$ and we obtain 
\begin{equation}\label{eq:CouplingLimit1}
\zeta 
=\frac{ C_g^2}{C_r(C_g + C_q) + C_g(C_g + NC_q)}< 1.
\end{equation}
This shows that independently of the circuit parameters the single-qubit ultrastrong coupling regime cannot be reached. 
Such a limit on the coupling parameter is consistent with more general no-go theorems discussed in Ref.~\cite{ViehmannPRL2009} or the absence of a SRT found in other explicit multi-qubit circuit QED models~\cite{Leib2012}. It can be traced back to the fact that in harmonic or weakly nonlinear systems the coupling $g\sim\sqrt{\bar \omega_q \bar \omega_r}$ is directly related to the qubit and the resonator frequency. To break this relation we now consider instead the charge qubit limit $E_J\ll E_C$ and $C_q V_G/(2e)=1/2$ [see Fig.~\ref{Fig1Setup} b)]. In this case the qubit states are superpositions of charge states, i.e. $|0\rangle=(|0e\rangle+|-2e\rangle)/\sqrt{2}$ and $|1\rangle=(|0e\rangle-|-2e\rangle)/\sqrt{2}$, where $|0e\rangle$ and $|-2e\rangle$ denote the states with zero and one excess Cooper pair on the island, respectively. Then $Q_0^q\approx e$ is approximately independent of $\bar \omega_q \simeq E_J$ and we obtain 
\begin{equation}\label{eq:CouplingLimit2}
\zeta=  
\dfrac{4C_g^2}{C_r(C_g + C_q) + C_g(C_g + NC_q)}\times \dfrac{E_C}{E_J}.
\end{equation}
Therefore, while in practice there might be additional constraints on the achievable parameters, there is no fundamental limit that prevents one from reaching the ultrastrong coupling regime even for a single qubit, for example, by simply lowering $E_J$. This is explicitly shown in Fig.~\ref{Fig1Setup} c) for a concrete set of realistic parameters and is in agreement with many previously analyzed circuits~\cite{BourassaPRA2009,NatafPRL2010,PeropadrePRL2013} based on the highly nonlinear flux qubit design~\cite{YouNature2011,FluxQubit}.

\subsection{Superradiant charge states}

\begin{figure}[t]
  \centering
    \includegraphics[width=0.48\textwidth]{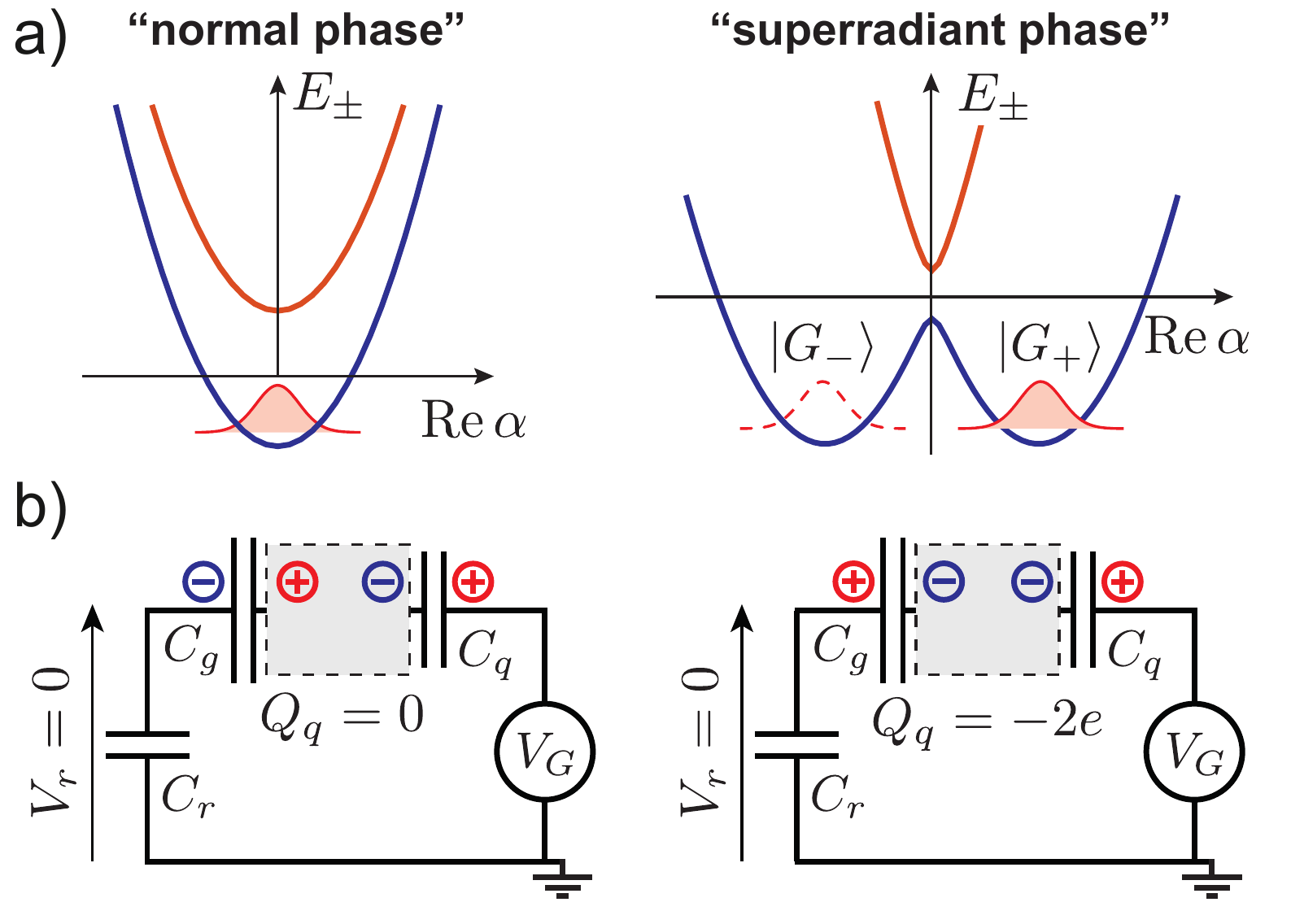}
      \caption{a) Plot of the two energy levels $E_\pm(\alpha)$ obtained from the diagonalization of the semiclassical qubit  Hamiltonian $H_{\rm sc}(\alpha)$ for $N=1$. b) Illustration of the classical charge configurations, which correspond to the two superradiant ground states $|G_\pm\rangle$. }
      \label{Fig2ChargeConfig1Qubit}
\end{figure}

Given the ability to reach the ultrastrong coupling regime $g>g_{\rm usc}$ for a single qubit, it is instructive to develop a physical picture for the ground state in this regime. To do so we consider a semiclassical model $H_{\rm sc}(\alpha)=\langle \alpha |H|\alpha\rangle$, where $a$ and $a^\dag$ are replaced by classical field amplitudes $\alpha$ and $\alpha^*$. For $N=1$ this model reads
\begin{equation}
H_{\rm sc}(\alpha)= \bar \omega_r |\alpha|^2  + \frac{\bar \omega_q}{2}\sigma_z + \frac{g}{2}(\alpha+\alpha^*) \sigma_x. 
\end{equation}
In Fig.~\ref{Fig2ChargeConfig1Qubit} a) we plot the two eigenenergies $E_\pm(\alpha)$ of $H_{\rm sc}(\alpha)$, which can be interpreted as Born-Oppenheimer potentials for a classical resonator with amplitude $\alpha$.  For $g=0$ the lowest potential curve is simply quadratic with a minimum at $\alpha=0$. This corresponds to the normal phase, where both the qubit and the resonator state are in the ground state.  For values $g\gg g_{\rm usc}$ the lowest potential curve exhibits two minima at $\alpha\approx \pm g/(2\bar \omega_r)$.  The two essentially degenerate ground states  $|G_\pm\rangle = |\pm\alpha\rangle |\mp  \rangle$, where $|\pm\rangle=(|0\rangle \pm |1\rangle)/\sqrt{2}$, then correspond to the superradiant states with a non-vanishing field expectation value.

The physical interpretation of the abstract states $|G_\pm\rangle$ is given in Fig.~\ref{Fig2ChargeConfig1Qubit} b), which shows the corresponding classical charge configurations in the limit where all the inductive energies can be neglected. In this limit the charge expectation values for these states are given by $\langle G_{\pm} | Q_r|G_\pm \rangle \simeq  \pm e $, $\langle G_-| Q_q| G_-\rangle \simeq 0$ and  $\langle G_+| Q_q| G_+\rangle \simeq -2e$. Most importantly, this simple electrostatic picture clearly illustrates that the main difference in the two configurations is the sign of the polarization charge on the coupling capacitance $C_g$. Although the system is in a superradiant state, the voltage across the resonator capacitance $C_r$ vanishes. Thus the meaning of a finite field expectation value $\langle a\rangle \neq 0$ in the ground state of a capacitively-coupled circuit QED system is that of a finite polarization charge, while the voltage state of the resonator is unchanged. A closely related observation about the difference between the electric field and the polarization field  has previously been pointed out in Ref.~\cite{Keeling2007} for atomic cavity QED systems, but it can be understood here in terms of even simpler electrostatic arguments.

\section{Ground states of the extended Dicke model}\label{sec:IDM}

Let us now return to the full circuit QED Hamiltonian~\eqref{eq:TLAHamiltonian}. Note that for $N\gg1$ neither Eq.~\eqref{eq:CouplingLimit1} nor Eq.~\eqref{eq:CouplingLimit2} prevents one from reaching the collective ultrastrong coupling regime, $N\zeta >1$, but now the additional qubit-qubit interactions $\sim D$ must be taken into account. Following the standard approach~\cite{Emary}, we first consider the limit $N\gg1$, and use the Holstein-Primakoff approximation to replace spins by harmonic operators, i.e.  $S_z\rightarrow b^\dag b-N/2$ and $S_x\rightarrow \sqrt{N}(b+b^\dag)/2$, where $[b,b^\dag]=1$. In this case we obtain the quadratic Hamiltonian (see App.~\ref{App:HP})
\begin{equation}\label{eq:Hlin}
H_{\rm HP} =  \bar \omega_r a^\dag a  + \bar \omega_q b^\dag b + G(a+a^\dag)(b + b^\dag) + D_N (b+b^\dag)^2,  
\end{equation}
where $G=g\sqrt{N}/2$ and $D_N=ND/4$. 
This Hamiltonian can be diagonalized by a Bogoliubov transformation~\cite{Emary,TodorovPRX2014,TodorovPRL2010}, from which we obtain 
the two eigenmode frequencies 
\begin{equation}\label{eq:Frequencies}
\omega_{\pm}^2=  \frac{1}{2}\bigg[  \bar \omega_r^2 + \Omega_q^2 \pm \sqrt{ (\bar \omega_r^2 - \Omega_q^2)^2 +16 G^2\bar \omega_r \bar{\omega}_q}\bigg],
\end{equation}
where $\Omega_q^2=\bar \omega_q(\bar \omega_q + 4 D_N)$. The SRT occurs, when the ground state of $H_{\rm HP}$ becomes unstable, i.e., when $\omega_-$ vanishes. This requires $G^2>\bar \omega_c \Omega_q/4$, or equivalently,  $N (g^2/\bar\omega_r)> ND +\bar \omega_q$, and for $D=0$ we recover the usual transition point mentioned above. However, in view of relation~\eqref{eq:D}, the excitation frequencies of the EDM remain positive for all parameter values. 
%
%
This means that in the present circuit QED setup a SRT in the conventional sense does not occur, and it is prevented by a mechanism, which is analogous to the effect of polarization interactions discussed, for example, in the context of  intersubband polariton systems~\cite{TodorovPRX2014,TodorovPRL2010}.
Since this effect is absent for a single qubit, there is a fundamental difference between single- and multi-qubit cavity QED settings, which does not follow from otherwise closely related studies of the $A^2$-term~\cite{Rzazewski,CiutiNatComm2010,ViehmannPRL2009}.

\begin{figure}[t]
  \centering
    \includegraphics[width=0.48\textwidth]{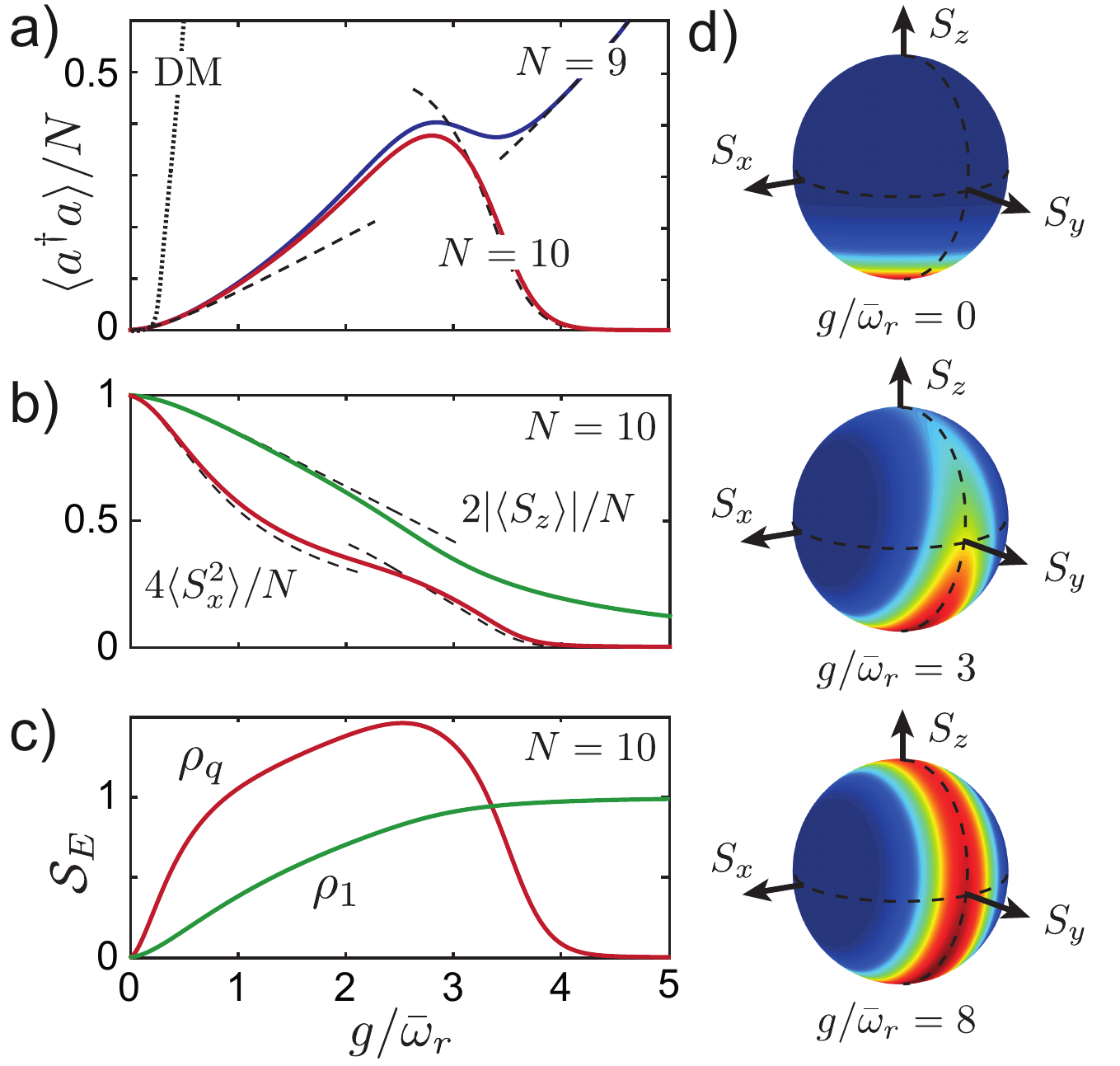}
      \caption{Ground-state expectation values of a) the photon number and b) the collective spin operators as a function of the coupling $g$. 
      The dotted line indicates the prediction from the Dicke model (DM) and the dashed lines the results obtained from $H_{\rm HP}$ in Eq.~\eqref{eq:Hlin} and $H_{\rm eff}$ in Eq.~\eqref{eq:Heff} in the weak and strong coupling limit, respectively.  c) Entanglement entropy $\mathcal{S}_E(\rho)=-{\rm Tr}\{ \rho \log_2(\rho)\}$ in the ground state $|G\rangle$ of Hamiltonian~\eqref{eq:TLAHamiltonian} evaluated for the reduced density matrix of the qubit subsystem, $\rho_q={\rm Tr}_{\rm r}\{|G\rangle\langle G|\}$, and for the reduced density matrix of a single qubit, $\rho_1={\rm Tr}_{N-1}\{\rho_q\}$.  d) Plot of the Q-function $Q(\vec n) =\langle \vec n| \rho_q |\vec n\rangle$, where $\vec n$ is a unit vector on the Bloch sphere and $|\vec n\rangle$ is the corresponding coherent spin state. Note that in b)-d) only the results for $N=10$ are shown and for all plots $\bar \omega_q/\bar \omega_r=0.5$ and $\delta=0$ have been assumed.}
      \label{Fig3GroundState}
\end{figure}

\subsection{Light-matter decoupling}

Having established the absence of a SRT in the linearized regime, we are now interested in the actual ground state of Hamiltonian~\eqref{eq:TLAHamiltonian} under the constraint $D=g^2/\bar \omega_r+\delta$, but for otherwise arbitrary coupling parameters. The inclusion of an additional offset $\delta\geq0$ is motivated by more general circuits discussed in Sec.~\ref{sec:FQ} below.
In Fig.~\ref{Fig3GroundState} a) and b) we plot  the expectation values of the mean photon number $\langle a^\dag a\rangle$ and the spin expectation values $\langle S_z\rangle $ and $\langle S_x^2\rangle$ as a function of the coupling $g$ and for $N=9$ and $N=10$ qubits. 
The plot shows a gradual increase of the photon number, which---as expected from the analysis above---varies smoothly across the SRT point $g=\sqrt{\bar \omega_r \bar \omega_q/N}$. This behavior can be fully understood from the linearized Hamilton $H_{\rm HP}$, from which we derive the approximate initial scaling (see App.~\ref{App:HP})
\begin{equation}\label{eq:NHP}
\langle a^\dag a\rangle \simeq \dfrac{Ng^2\bar{\omega}_q}{4(\bar{\omega}_r + \bar{\omega}_q)^2[\bar{\omega}_q + N(D -g^2/\bar\omega_r)]}.
\end{equation}
The validity of this results requires a low excited-state population of each individual qubit, i.e. $g\ll \bar \omega_r+\bar \omega_q$. Beyond this point nonlinear effects start to play a significant role and surprisingly, for an even number of qubits we observe a sudden decrease of the photon number and $\lim_{g\rightarrow \infty}\langle a^\dag a\rangle=0$. At the same time the qubits remain in a highly excited state, i.e. $\langle S_z\rangle\approx 0$, with a vanishing polarization along $x$, i.e. $\langle S_x\rangle, \langle S_x^2\rangle \rightarrow 0$.  In Fig.~\ref{Fig3GroundState} c) we also plot the entanglement entropy, $\mathcal{S}_E(\rho)= - {\rm Tr}\{\rho \log_2(\rho)\}$, for the reduced qubit density operator and for the density operator of a single qubit. It shows that while the spin and cavity subsystems become decoupled at large $g$, the qubits remain highly entangled among themselves.   

We remark that other light-matter decoupling mechanisms have recently been described in extended multi-mode systems~\cite{DeLiberato,GarciaRipoll2015,Malekakhlagh2015}. There the inclusion of $A^2$-like terms expels the relevant field modes from the coupling region, an effect which can already be described within a linearized model.  In the present single-mode setup such a mechanism is not possible, and the above observations already indicate that here the field decoupling is a highly nonlinear and nonclassical effect. A heuristic explanation for this behavior can be obtained by considering the limit $D\gg g, \bar \omega_{r,q}$, where the qubit-qubit interaction is the dominant energy and therefore favors the state $|m_x=0\rangle$, where $S_x|m_x\rangle=m_x |m_x\rangle$,  as the ground state. Since the coupling to the field $\sim S_x$, it then also vanishes and the resonator mode decouples. This is visualized in Fig.~\ref{Fig3GroundState} d), in terms of the $Q$-function of the reduced qubit state on the Bloch sphere, which approaches a circle in the $x=0$ plane  for large $g$. For odd $N$, there is no $m_x=0$ state, which explains why this decoupling mechanism does not take place for the example of $N=9$ qubits.

\subsection{Effective low energy model}

\begin{figure}[t]
  \centering
    \includegraphics[width=0.48\textwidth]{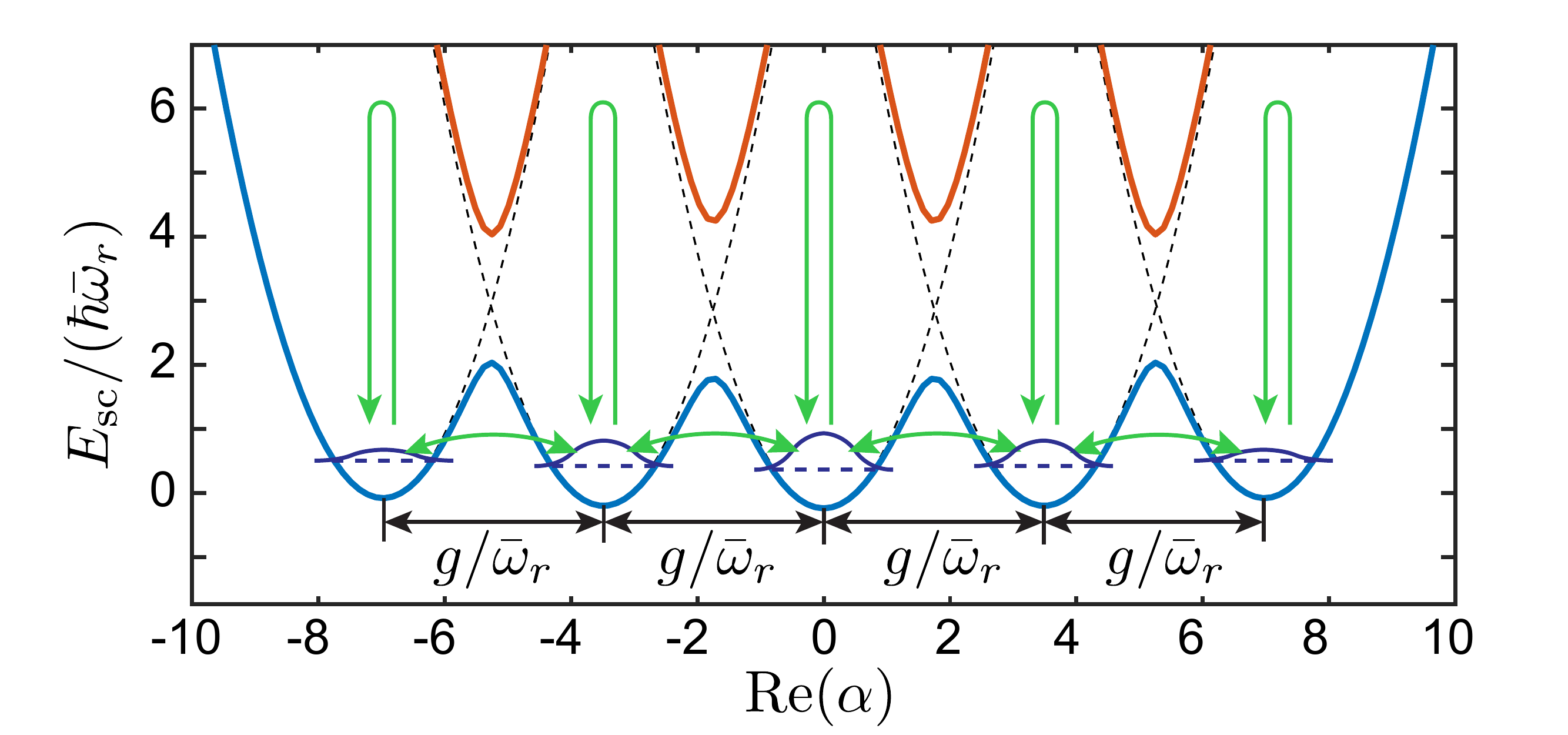}
      \caption{Plot of the lowest two energy levels $E_{\rm sc}$ obtained from the diagonlization of the semiclassical qubit  Hamiltonian $H_{\rm sc}(\alpha)=\langle \alpha |H|\alpha\rangle$, and for $g/\bar \omega_r=3.5$, $\bar \omega_q/\bar \omega_r=1$ and $N=4$. The arrows indicate first and second order processes induced by $H_1=\bar \omega_q S_z$ on the otherwise degenerate ground state manifold.}
      \label{Fig4BOPotentials}
\end{figure}

\begin{figure}[t]
  \centering
    \includegraphics[width=0.48\textwidth]{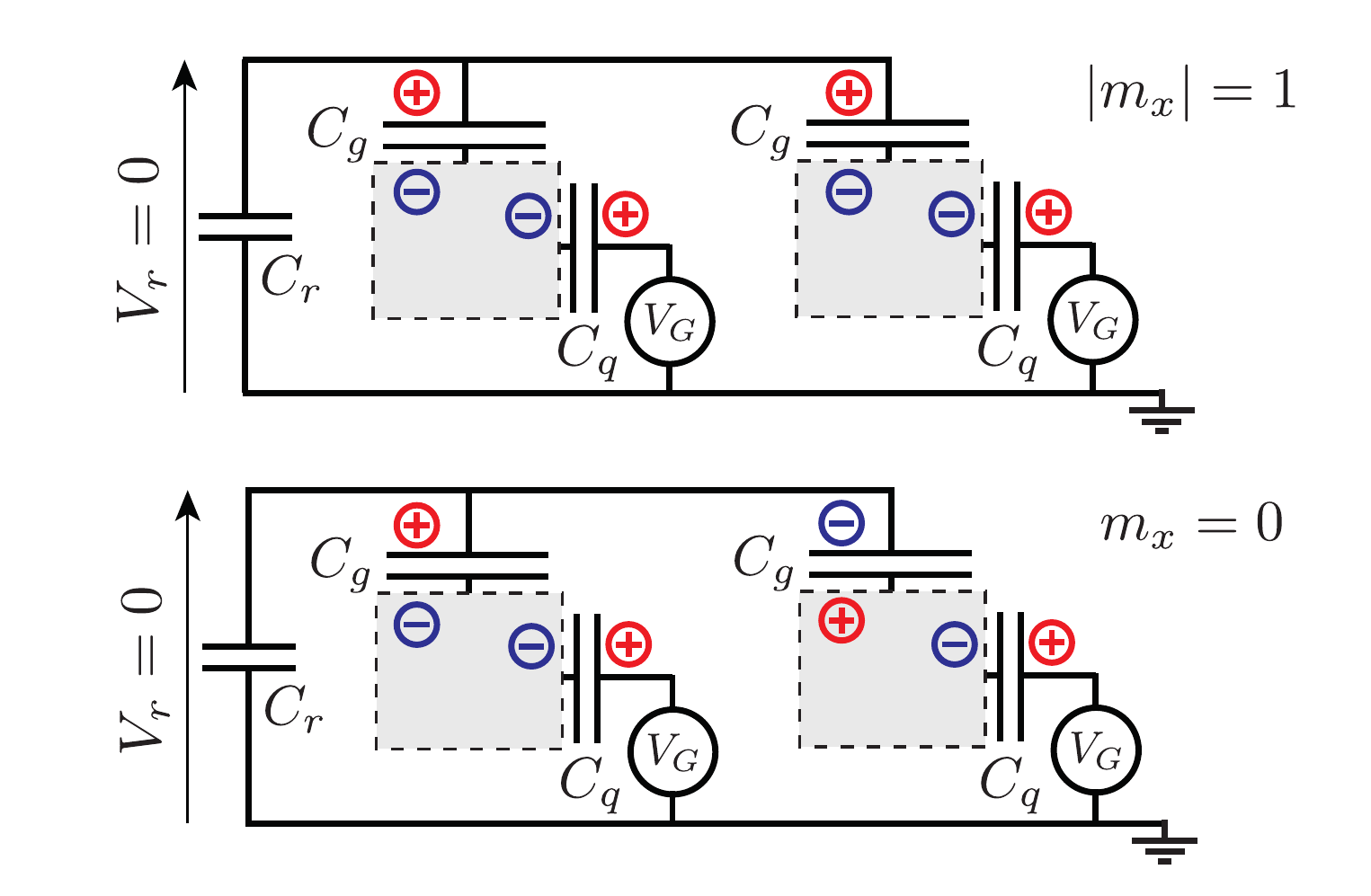}
      \caption{Illustration of the classical charge configurations, which correspond to two different qubit states with different values of $m_x$. In both cases the capacitive energies are the same, which when generalized to $N$ qubits explains the existence of $2^N$-fold degenerate manifolds in the limit $g\rightarrow \infty$. As in the single qubit case, the average voltage across the resonator capacitance $C_r$ vanishes for all those states. }
      \label{Fig5ChargeConfig2Qubits}
\end{figure}

Let us now develop a more accurate description of the system in the regime $g/\bar \omega_r> 1$. We write $H=H_0+H_1$, where $H_1=\bar \omega_q S_z$. The first term, $H_0$, commutes with $S_x$ and therefore it can be diagonalized exactly, $H_0|\Psi_{s,m_x,n}\rangle = E_{m_x,n}|\Psi_{s,m_x,n}\rangle$ (see App.~\ref{App:USCPerturbationTheory}).  The eigenstates are given by
\begin{equation}\label{eq:Eigenstates}
|\Psi_{s,m_x,n}\rangle =   e^{-\frac{g}{\bar \omega_r}Ê(a^\dag- a) S_x} |s,m_x\rangle |n\rangle,
\end{equation} 
where $|n\rangle$ are photon number states and $|s,m_x\rangle$ are collective spin states of given total spin $s=0,\dots,N/2$ and projection quantum number $m_x=-s,\dots,s$. Therefore, the eigenstates of $H_0$ are simply harmonic oscillator states, which are displaced by an amount $-m_x g/\bar \omega_r$. The corresponding eigenenergies are 
\begin{equation}
E_{m_x,n}= \delta m_x^2 + \bar \omega_r n,
\end{equation}
which for $\delta\rightarrow 0$ become independent of the spin quantum numbers. This means that in this limit each $n$-manifold contains a set of $2^N$ degenerate qubit states. The energy penalty from the $S_x^2$-term is exactly compensated by a lowering of the interaction energy when the resonator mode is displaced. This is illustrated in Fig.~\ref{Fig4BOPotentials}, where we plot the lowest two eigenvalues of the semiclassical qubit Hamiltonian, $H_{\rm sc}(\alpha)$, for $s=N/2$. 
Compared to the double-well potential structure for the single qubit case, the lowest Born-Oppenheimer potential now displays a multi-well potential landscape with $(2s+1)$ nearly degenerate minima. Again this large degree of degeneracy can be understood from the corresponding classical charge configurations, as illustrated in Fig.~\ref{Fig5ChargeConfig2Qubits} for the case $N=2$. For both qubit configurations, which correspond to different values of $m_x$ the electrostatic energy is concentrated only in the gate and coupling capacitances $C_q$ and $C_g$, and it is therefore identical. Note that the DM with $D=0$ would energetically favor states with  $|m_x|=1$ over a $m_x=0$ state and lead to inconsistencies with basic electrostatic considerations.

To model the low energy properties of $H$ we now focus on the ground state manifold, $n=0$, and include quantum corrections due to $H_1$. This term first of all couples neighboring $m_x$ levels within this manifold, i.e., $\langle  \Psi_{s,m_x,0}|S_z  |\Psi_{s,m_x\pm1,0}\rangle = e^{-g^2/2\bar \omega_r^2} \langle s, m_x|S_z  |s, m_x\pm 1\rangle$. In addition, $H_1$ also couples to states in energe-tically higher manifolds, $n> 0$. This effect can be treated in second order perturbation theory as detailed in App.~\ref{App:USCPerturbationTheory}. These second order corrections predominately lead to an energy shift $\sim [m_x^2-s(s+1)] \bar \omega_q^2 \bar \omega_r/(2g^2)$, which singles out the $s=N/2$ states as the lowest energy manifold. By defining operators $\tilde S_k$ via the relation $\langle \Psi_{s, m_x^\prime,0}|\tilde S_k |\Psi_{s, m_x,0}\rangle = \langle s, m_x^\prime|S_k |s, m_x\rangle$, we obtain within this manifold the effective collective spin Hamiltonian
 \begin{equation}\label{eq:Heff}
 \begin{split}
 H_{\rm eff} \simeq  & \left( \delta + \frac{\bar \omega_q^2 \bar \omega_r}{2g^2} \right) \tilde S_x^2 +  \bar \omega_q e^{-g^2/2\bar \omega_r^2}\tilde S_z.
 \end{split}
 \end{equation}
From this  model we immediately see that for $g/\bar \omega_r\gg 1 $ the coupling to neighboring $m_x$ states is exponentially suppressed and for even $N$ the ground state is indeed $|G\rangle=|\Psi_{N/2,m_x=0,n=0}\rangle$ with a vanishing photon number. By taking into account first order corrections from states $|\Psi_{N/2,m_x=\pm1,n=0}\rangle$ we obtain the approximate scalings
\begin{equation}
\begin{split}
\langle a^\dag a\rangle \approx \dfrac{N(N + 2)g^6}{2\bar{\omega}_r^4\bar{\omega}_q^2} e^{-g^2/\bar{\omega}_r^2},\qquad \Delta E \approx \dfrac{\bar{\omega}_r\bar{\omega}_q^2}{2g^2},
\end{split}
\end{equation} 
where $\Delta E=E_1-E_0$ is the gap between the ground and the first excited state. Note that this non-exponential closing of the energy gap is very atypical for qubit-resonator models in the ultrastrong coupling regime. For an odd number of qubits there is no $m_x=0$ state and the ground state is $|G\rangle \simeq (|\Psi_{N/2,m_x=-1/2,n=0}\rangle+|\Psi_{N/2,m_x=1/2,n=0}\rangle)/\sqrt{2}$ and $\langle a^\dag a\rangle \simeq g^2/(4\bar \omega_r^2)$ simply increases in the large $g$ limit. The next higher state is the corresponding antisymmetric superposition and therefore the energy splitting now exhibits the usual exponential scaling, i.e., $\Delta E\simeq (N + 1)\bar \omega_q e^{-g^2/2\bar \omega_r^2}$. Although for certain quantities the admixture of higher $n$-levels must be taken into account (see App.~\ref{App:USCPerturbationTheory}), we find that within its range of validity, $\sqrt{s(s+1)} <g^2/(\bar \omega_r \bar \omega_q) $ and $g/\bar \omega_r>1$, the effective model $H_{\rm eff}$ provides an accurate description of the low energy properties of an ultrastrongly coupled collective circuit QED system.

\subsection{Energy spectrum}
\begin{figure}[t]
  \centering
    \includegraphics[width=0.48\textwidth]{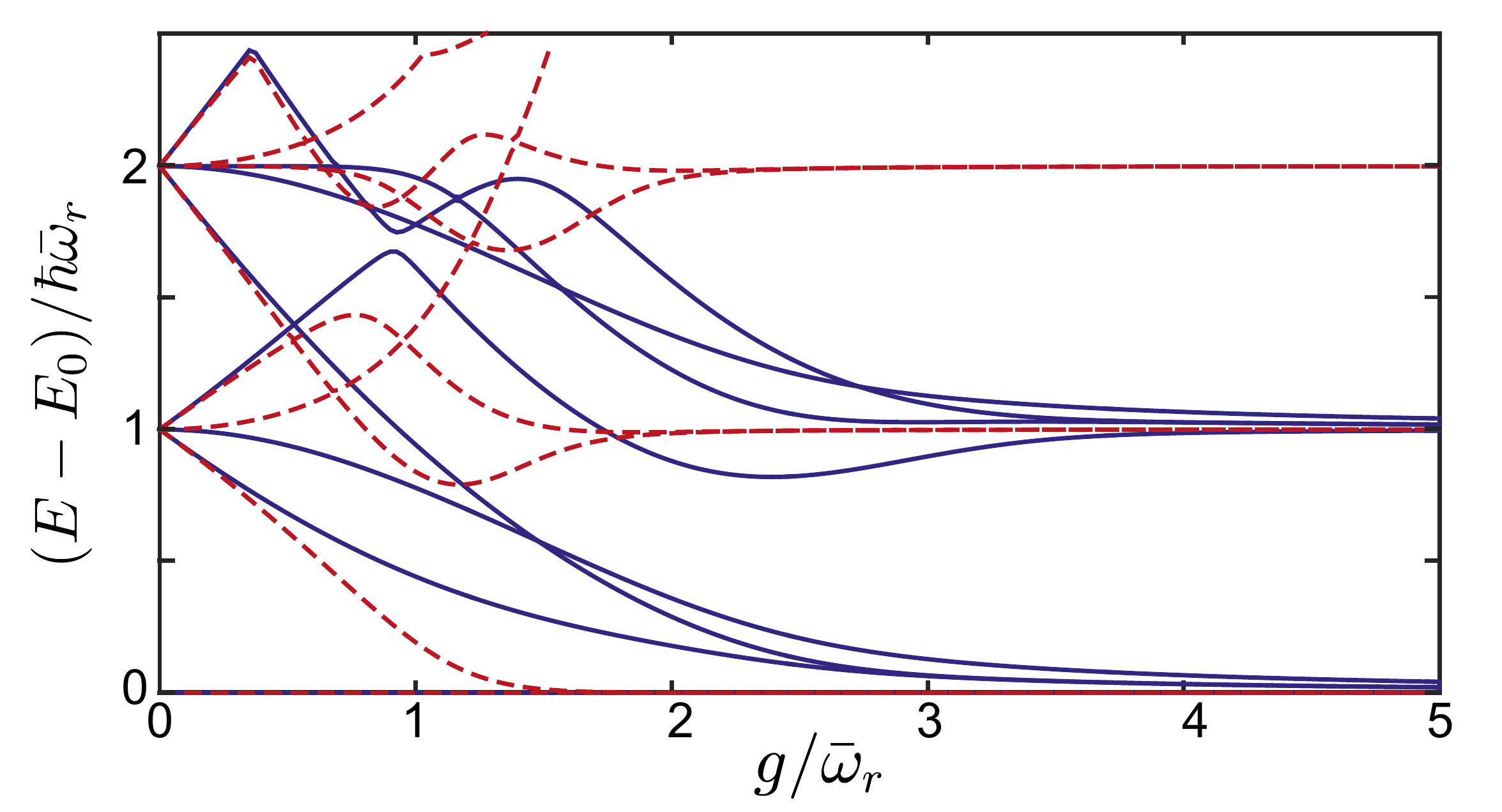}
      \caption{The lowest excitation energies $\Delta E_n=E_n-E_0$ of the EDM~\eqref{eq:TLAHamiltonian} for $N=2$ qubits are plotted as a function of the coupling strength $g$ and for $\bar \omega_r=\bar \omega_q$. The dashed lines show the corresponding excitation energies for the standard DM, where $D=0$.  }
      \label{Fig6Spectrum}
\end{figure}

Apart from the above described changes of the ground state properties, the appearance of the $DS_x^2$ term will also change the excitation spectrum of multi-qubit circuit QED systems in the ultrastrong coupling regime. This is exemplified in Fig.~\ref{Fig6Spectrum} for the case $N=2$, where the lowest few excitation energies $\Delta E_n=E_n-E_0$ are plotted as function of the coupling $g$ and compared with the corresponding results for the DM. One clearly sees the emergence of the $2^N$-fold degenerate manifolds for large $g$, which already for two qubits is in clear contrast to the two-fold degenerate energy manifolds known from the DM. Spectroscopically, the difference between the two models becomes most significant in the region $1<g/\bar \omega_r <2$, where the crossover from the strong to the ultrastrong coupling regime takes place.

\section{Flux qubit circuits and experimental implementations}\label{sec:FQ}

\begin{figure}[t]
  \centering
    \includegraphics[width=0.48\textwidth]{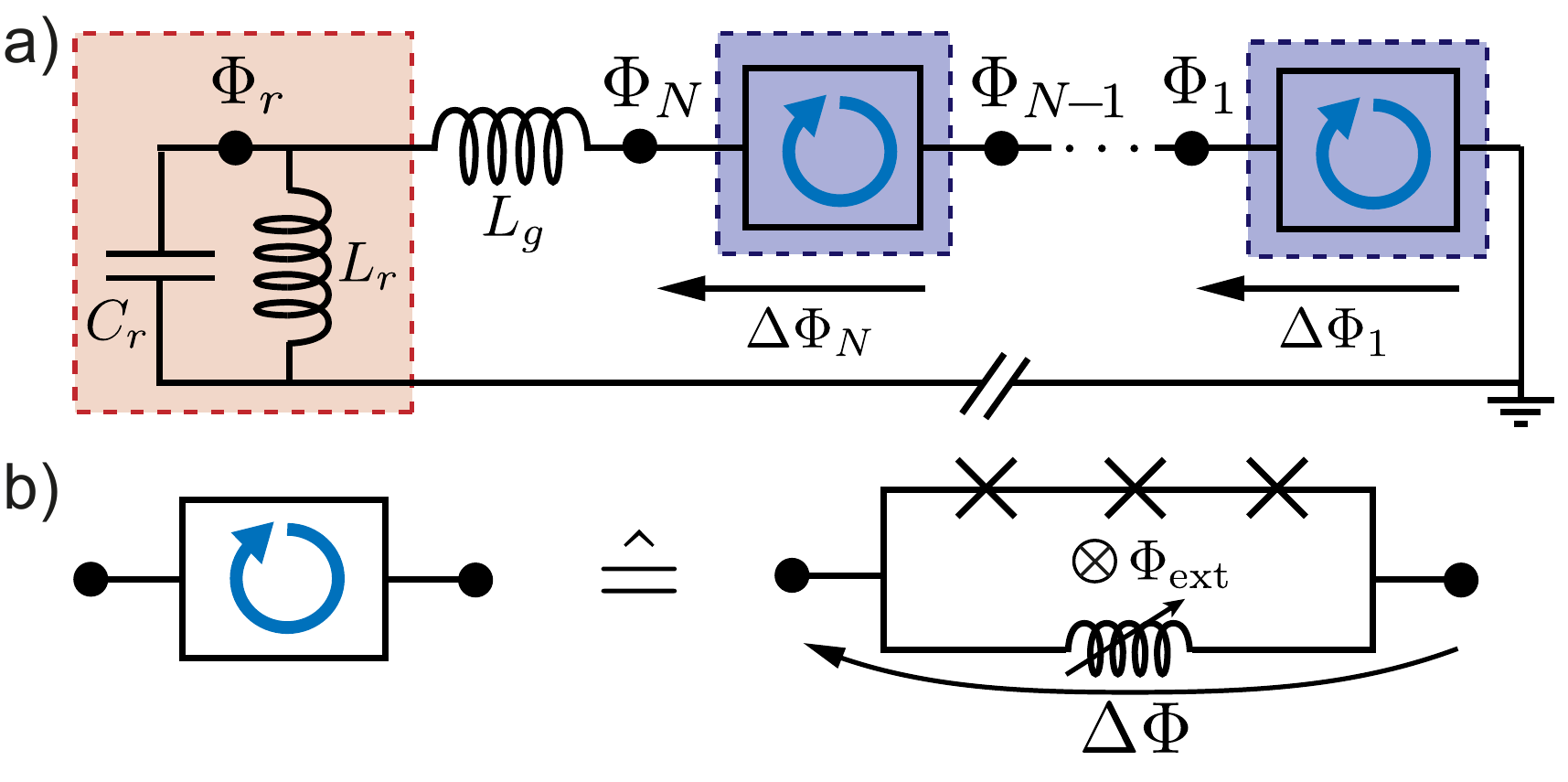}
      \caption{Circuit model for a collective QED system with inductively coupled flux qubits. When expressed in terms of the flux across each qubit, $\Delta \Phi_i=\Phi_i-\Phi_{i-1}$, the magnetic energy $(\Phi_r- \sum \Delta\Phi_i)^2/(2L_g)$ associated with the inductance $L_g$ leads to qubit-resonator as well as qubit-qubit interactions, with the relation $D=g^2/\bar \omega_r+\delta$, where $\delta>0$. b)  Specific realization of a flux qubit based on the design used in Refs.~\cite{Yoshihara2016,FornDiaz2016} to reach the ultrastrong coupling regime.}
      \label{Fig7FluxQubit}
\end{figure}

The results presented so far have been explicitly derived for a charge coupled circuit, but the appearance of a $DS_x^2$ term in multi-qubit circuits is a more general phenomenon. In Fig.~\ref{Fig7FluxQubit} a) we show an equivalent circuit model for $N$ flux qubits coupled inductively to a single $LC$ resonator. The Lagrangian for this circuit is 
\begin{equation}\label{eq:LagrangianFQ}
\mathcal{L}=\frac{C_r\dot \Phi_r^2}{2} - \frac{\Phi_r^2}{2L_r} - \frac{(\Phi_r-\Phi_N)^2}{2L_g}+ \sum_{i=1}^N   \mathcal{L}_q(\Delta \Phi_i,\Delta \dot \Phi_i).
\end{equation}
Here  $\mathcal{L}_q$ denotes the Lagrangian for a single flux qubit, which is a function of the phase difference $\Delta \Phi_i=\Phi_i-\Phi_{i-1}$, and depending on the exact design, of additional local degrees of freedom~\cite{BourassaPRA2009,NatafPRL2010,PeropadrePRL2013,YouNature2011,FluxQubit}. Since for this circuit there are no capacitive couplings between the individual components, the derivation of the corresponding circuit Hamiltonian can be performed in a straight forward manner. In particular, by writing $\Phi_N=\sum_i \Delta\Phi_i$, one immediately sees that the resulting inductive interaction between $\Phi_r$ and $\Phi_N$ can be grouped as
\begin{equation}
\begin{split}
 \frac{(\Phi_r- \Phi_N)^2}{2L_g}
=  \frac{\Phi_r^2}{2L_g} -\sum_i   \frac{\Delta\Phi_i\Phi_r }{L_g}  +  \sum_{i,j} \frac{\Delta\Phi_i\Delta\Phi_j }{2L_g}.
\end{split}
\end{equation}
The first $A^2$-like term leads to a renormalization of the resonator inductance $L_r\rightarrow \bar L_r =(L_rL_g)/(L_r+L_g)$, which however does not scale with $N$.  The other two terms represent the collective qubit-photon coupling and the collective qubit-qubit interactions, respectively. By writing $\Phi_r=\sqrt{\hbar /2 C_r \bar \omega_r}(a^\dag+a)$ and within the validity of the two-level approximation, i.e. $\Delta \Phi_i = \Phi_q^0 \sigma_x^i$, we recover the EDM~\eqref{eq:TLAHamiltonian} with the more general relation
\begin{equation}
D= \frac{g^2}{\bar \omega_r}\left(1+\frac{L_g}{L_r}\right) = \frac{g^2}{\bar \omega_r}+\delta,
\end{equation}
where $\delta>0$, as assumed in the analysis above.

The flux qubit circuit shown in Fig.~\ref{Fig7FluxQubit} a) is also very promising for first experimental realizations of the described models. Compared to capacitive interactions, where $g/\bar \omega_r\sim \sqrt{\bar Z_r/R_K}$  ($R_K=h/(e^2)$ is the quantum resistance), the inductive coupling $g/\bar \omega_r\sim \sqrt{R_K/\bar Z_r}$ scales more favorable with the resonator impedance $\bar Z_r=\bar L_r/\bar C_r$ \cite{DevoretAnnPhys2007}. Fig.~\ref{Fig7FluxQubit} b) shows a sketch of a three-junction flux qubit with an additional tunable inductance. This design has been used in recent experiments to demonstrate ultrastrong coupling conditions in a single mode setup~\cite{Yoshihara2016} as well as for an open transmission line~\cite{FornDiaz2016} and exhibits a large degree of tuneability. The addition of a second qubit in such setups, would already allow the observation of the described decoupling effect, by either looking at the excitation spectrum shown in Fig.~\ref{Fig6Spectrum}, or by measuring the predicted anti-correlations of the qubit flux-states in the ground state of the system.

\subsection{Other experimental considerations}

\begin{figure}[t]
  \centering
    \includegraphics[width=0.48\textwidth]{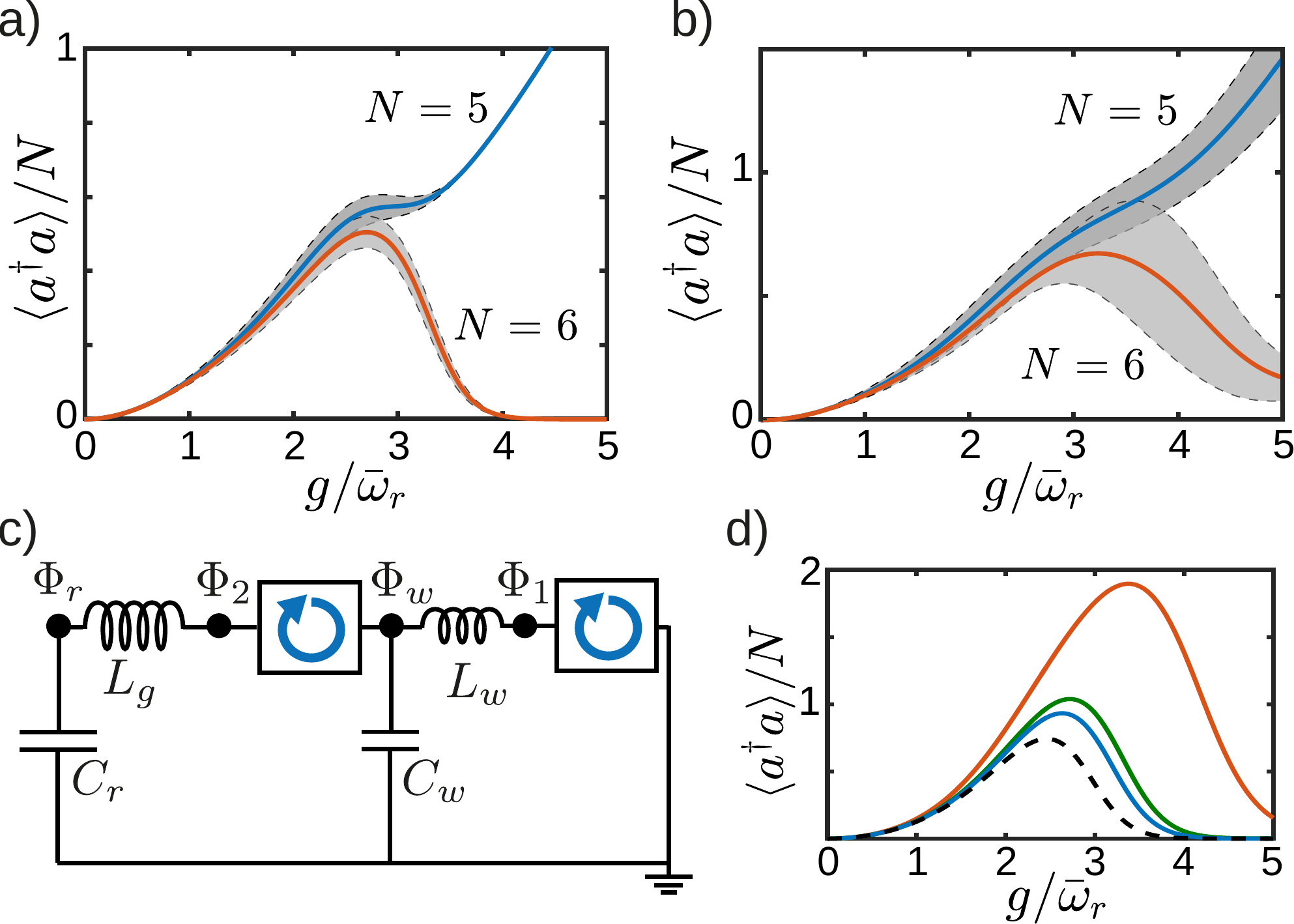}
      \caption{a) The ground-state photon number $\langle a^\dag a\rangle$ is plotted for disordered system, where the qubit frequencies $\bar \omega_q^i$ are randomly chosen from a uniform distribution in the interval $[\bar \omega_q-0.5 \bar \omega_q,\bar \omega_q+0.5 \bar \omega_q]$. b) The same as in a), but with a random distribution of coupling constants $g_i$ chosen from the interval $[g-0.3 g,g+0.3 g]$. The other parameters for this plot are the same as in Fig.~\ref{Fig3GroundState}. c) Circuit model for two flux-qubits, which are connected via a wire of finite inductance $L_w$ and capacitance $C_w$. The frequencies $\bar \omega_r$ and $\omega_{\rm ex}$ are identified with the lower and higher resonator frequencies and $g$ is identified with the coupling of the first qubit with the lower resonator mode. Other coupling constants are identified as $ g_{ik} $ where $ i = 1, 2 $ stands for the qubits and $ k = 1, 2 $ for the modes, $ k = 1 $ is the lower and $ k = 2 $ the higher. Thus, with this naming convention $ g \equiv g_{11} $. All of the parameters can be derived from the circuit by diagonalizing the harmonic part of the circuit's Hamiltonian. d) Ground-state photon number in the presence of a higher mode. For the plot we use $ L_w = L_g/3 $, $ C_w = C_r/87 $, $ \omega_{\rm ex}/\bar{\omega}_r = 21.56 $, $ g_{12}/g = 2.69 $, $ g_{21}/g = 1.00 $, $ g_{22}/g = -8.04 $ (blue), $ L_w = L_g $, $ C_w = C_r/29 $, $ \omega_{\rm ex}/\bar{\omega}_r = 10.86 $, $ g_{12}/g = 3.35 $, $ g_{21}/g = 1.02 $, $ g_{22}/g = -3.30 $ (green) and $ L_w = 3L_g $, $ C_w = 3C_r/29 $, $ \omega_{\rm ex}/\bar{\omega}_r = 7.61 $, $ g_{12}/g = 5.15 $, $ g_{21}/g = 1.08$, $ g_{22}/g = -1.59 $ (red). For all curves we have chosen $ L_g = 1.5\,\mathrm{nH} $ and $ C_r = 0.25\,\mathrm{pF} $. Dotted line displays the single-mode result. The values for $L_w$ and $C_w$ in the three curves  correspond approximately to a $ \sim 200\,\mathrm{\mu m} $ (blue), $ \sim 500\,\mathrm{\mu m} $ (green) and $ \sim 1\,\mathrm{mm} $ (red) long wire between the qubits. For all curves the qubit parameters have been adjusted to give the same frequency $\bar \omega_q = \bar \omega_r/2$.}
      \label{Fig8Experiments}
\end{figure}

In our analysis so far we have assumed identical couplings $g_i=g$ and identical qubit frequencies $\bar \omega_q^i=\bar \omega_q$, which can be hard to realize in practice~\cite{Macha2014,Kakuyanagi2016}. However, in Fig.~\ref{Fig8Experiments} we re-evaluate the light-matter decoupling effect for a disordered system, where we allow for individual variations of the $g_i$ and $\bar \omega_q^i$. We see that even at a very high level of disorder there are small quantitative differences, but almost no qualitative changes of the predicted features. Surprisingly, even the parity oscillations, which we explained above in terms of a fully symmetric coupling, are very robust with respect to variations in the coupling constants and all curves are still clearly distinct from the sharp increase of $\langle a^\dag a\rangle$ expected from the DM. Therefore, the predicted effects should be observable even in systems with only a limited amount of tuneability. 

Finally, our model assumes the coupling of all qubits to a single resonator mode. This is in general a good approximation for lumped element resonators, where the fundamental electromagnetic mode can be well separated from all higher excitations. For example, considering a typical $LC$ resonator of spatial extent $d\sim 500\,\mu$m and a fundamental mode of $\bar \omega_r/(2\pi) \sim 5$ GHz~\cite{FornDiazPRL2010,Yoshihara2016} one expects higher order modes at a frequency $\omega_{\rm ex}/(2\pi)\sim c/(6d)\sim 100$ GHz~\cite{PrivComm}. At the same time such a circuit can incorporate tens of flux qubits of size $\sim 10\,\mu$m. To understand the validity of the single mode approximation more quantitatively, we investigate a two mode setup shown in Fig.~\ref{Fig8Experiments} c), where the ideal wire connecting two flux qubits is replaced by a finite inductance $L_w$ and capacitance $C_w$. This circuit is used to model an higher excited mode with frequency $\omega_{\rm ex}\sim 1/\sqrt{L_w C_w}$. Fig.~\ref{Fig8Experiments} d) shows the dependence of the photon number $\langle a^\dag a\rangle$ in the ground state for varying ratios $\omega_{\rm ex}/\bar \omega_{r}$. Again we observe a robustness of the decoupling effect and for experimentally relevant regime $\omega_{\rm ex}/\bar \omega_{r}\geq 20$ (approximately corresponding to a $\sim 200\,\mu$m long wire) there is no significant influence of the higher mode. This ability to realize a single mode setup in the ultrastrong coupling is also one of the key advantages of circuit QED, since such a separation of modes cannot be achieved in the optical regime.

\section{Conclusions}\label{sec:Conclusions}

In summary we have analyzed collective interactions in circuit QED systems in the ultrastrong coupling regime. Going beyond the previously discussed $A^2$-corrections, we have identified the important role of qubit-qubit interactions, which generically appear in the fundamental description of such circuits. In particular, we have shown that the ground state of the resulting EDM exhibits many features that are exactly opposite to what is expected from the often considered DM physics. Apart from the absence of a superradiant phase, this includes light-matter decoupling at very strong interactions, a high degree of entanglement between the qubits and the existence of manifolds with an exponentially large number of nearly degenerate states. These predictions can already be tested with a minimal setup consisting of two flux qubits coupled to a lumped element resonator, similar to existing experimental setups~\cite{Yoshihara2016,FornDiaz2016}.

On a broader scope, we have presented in this work a fully microscopic derivation of a minimal model describing multiple (artificial) atoms coupled to a single radiation mode, where compared to optical systems the single mode or two level approximations can be rigorously justified by an appropriate circuit design. The resulting model is identical to the single-mode Hopfield model, which is usually derived for macroscopic dielectrics and has been successfully applied to describe collective ultrastrong coupling effects in solid-state cavity QED systems~\cite{TodorovPRL2010}. Our analysis reveals for the first time the highly nontrivial quantum mechanical features of this fundamental model in the regime  where each individual atom is coupled ultrastrongly to the radiation field.

\acknowledgements
We thank J. Majer and S. Putz for stimulating discussions. 
This work was supported by the Austrian Science Fund (FWF) through SFB FOQUS F40, DK CoQuS W 1210 and the START grant Y 591-N16 and by the European Commission through the FP7/ITC Project SIQS (600645) and the Marie Sklodowska-Curie
Grant IF 657788. J.J.G.R. acknowledges support from the Spanish Mineco Project FIS2012-33022 and the CAM Research Network QUITEMAD+.

\appendix

\section{Circuit QED Hamiltonian}\label{App:CircuitHamiltonian}
In this Appendix we provide additional details on the derivation of the charge Hamiltonian~\eqref{eq:FullHamiltonian}. To perform the Legendre transformation we write the circuit Lagrangian~\eqref{eq:Lagrangian} as
\begin{align}
\mathcal{L} = \dfrac{1}{2}\dot{\bm{\Phi}}^{\rm T}\mathcal{C}\dot{\bm{\Phi}} - \dot{\bm{\Phi}}^{\rm T}\bm{a} + \dfrac{NC_qV_G^2}{2} - V_{\rm pot}(\{ \Phi_\eta \}),
\end{align}
where $ \dot{\bm{\Phi}} = ( \dot{\Phi}_r, \dot{\Phi}_1, \dots, \dot{\Phi}_N)^{\rm T} $, $V_{\rm pot}(\{ \Phi_\eta \})$ is the potential energy and $\bm{a} = \left( 0, C_qV_G, \dots, C_qV_G \right)^{\rm T} $. For the circuit shown in Fig.~\ref{Fig1Setup} a) the capacitance matrix is \begin{align}
    \mathcal{C} =
    \begin{pmatrix}
        C_r + NC_g & -C_g & -C_g & \dots & -C_g\\
        -C_g & C_q + C_g & 0 & \dots & 0\\
        -C_g & 0 & C_q + C_g & \dots & 0\\
        \vdots & \vdots & \vdots & \ddots & \vdots\\
        -C_g & 0 & 0 & \dots & C_q + C_g\\
    \end{pmatrix}.
\end{align}
The Hamiltonian function is obtained from the Lagrangian via a Legendre transformation and can be written in vector notation as
\begin{align}\label{eq:Hamiltonianfunction}
    \mathcal{H} = \dfrac{1}{2}\left( \bm{Q} + \bm{a} \right)^{\rm T} \mathcal{C}^{-1} \left( \bm{Q} + \bm{a} \right) + V_{\rm pot}(\{ \Phi_\eta \}),
\end{align}
where $ \bm{Q} = \left( Q_r, Q_1, \dots, Q_N \right)^T $ is the vector of conjugate charges $\bm{Q} = \partial\mathcal{L}/\partial\bm{\dot{\Phi}} $. The inverse of the capacitance matrix is of the form 
\begin{equation}\label{eq:Cinv}
 \frac{1}{\mathcal{C}}=\frac{1}{\bar C^2}
\left(\begin{array}{ccccc}
C_q + C_g  &   C_g & C_g &\dots & C_g           \\
C_g        & X     &Y &\dots &Y \\
C_g       &Y   & X  &\dots  &Y  \\
\vdots & \vdots & \vdots & \ddots &\vdots  \\
C_g         &Y &Y  &\dots &X  \\
\end{array}\right),
\end{equation}
where $\bar C^2= C_rC_q + C_g\left( C_r + NC_q \right)$ and we used the abbreviations $ X = C_r+C_g + (N-1)C_gC_q/(C_g+C_q) $, $ Y = C_g^2/(C_g+C_q) $. The result for the charge Hamiltonian~\eqref{eq:FullHamiltonian} then follows directly from Eqs.~\eqref{eq:Hamiltonianfunction} and~\eqref{eq:Cinv}.

\section{Holstein-Primakoff Approximation}\label{App:HP}
The Holstein-Primakoff approximation is based on an exact mapping of collective spin operators $S_\pm=S_x\pm iS_y$ and $S_z$  onto a bosonic mode with annihilation operator $b$~\cite{HolsteinPrimakoff,Emary},
\begin{align}
S_+ =& b^\dag \sqrt{ÊN -b^\dag b},  \\
S_- =& \sqrt{ÊN -b^\dag b} \,b,  \\
S_z=&\left( b^\dag b -\frac{N}{2}\right).
\end{align} 
Under the assumption that $N\gg1$ and the total number of excitations remains small, i.e. $\langle b^\dag b\rangle/N\ll 1$ we can approximate $S_x \simeq \sqrt{N}(b^{\dagger} + b)/2$ in the EDM~\eqref{eq:TLAHamiltonian}, and we obtain the quadratic Hamiltonian $H_{\rm HP}$ given in Eq.~\eqref{eq:Hlin}.
This Hamiltonian can be diagonalized and written as~\cite{Emary}
\begin{align}
    H_{\rm HP} = \sum_{\alpha = \pm} \hbar\omega_{\alpha}c^{\dagger}_{\alpha}c_{\alpha},
\end{align}
where the excitation frequencies are given by
\begin{align}
    \omega_{\pm}^2 = \dfrac{1}{2} \left[ \bar{\omega}_r^2 + \Omega_q^2 \pm \sqrt{\left( \bar{\omega}_r^2 - \Omega_q^2 \right)^2 + 16G^2\bar{\omega}_r\bar{\omega}_q} \right]
\end{align}
and $ \Omega_q^2 = \bar{\omega}_q \left( \bar{\omega}_q + 4D_N \right) $. By using the relation $D=g^2/\bar \omega_r+\delta$ we find that both excitation frequencies are positive as long as $\delta\geq 0$ and $\bar \omega_q>0$.

%

The ground state $|G\rangle $ of this Hamiltonian is characterized by $ c_{\pm}| G \rangle = 0 $. Using this property we can calculate the average photon number in the ground state. We obtain
\begin{align}
    \langle a^{\dagger}a \rangle = \dfrac{\cos \left( 2\theta \right) A_- + A_+ - 4}{8},
\end{align}
where we used the short-hand notation
\begin{align}
    \cos \left( 2\theta \right) &= \dfrac{\bar{\omega}_r^2 - \Omega_q^2}{\sqrt{\left( \bar{\omega}_r^2 - \Omega_q^2 \right)^2 + 16G^2\bar{\omega}_c\bar{\omega}_q}},\\
    A_\pm  &= \dfrac{\left( \omega_+\omega_- \pm \bar{\omega}_{r}^2 \right) \left( \omega_+ \pm \omega_- \right)}{\bar{\omega}_{r}\omega_+\omega_-}.
\end{align}
For small $g$ the result for $\langle a^\dag a\rangle$ can be further simplified to the expression given in Eq.~\eqref{eq:NHP} in the main text.

\section{Ultrastrong coupling perturbation theory}\label{App:USCPerturbationTheory}
In this Appendix we summarize the details of the perturbation theory used to describe the ground state properties in the ultrastrong coupling regime $g>\bar \omega_r,\bar \omega_q$. In this regime the full Hamiltonian can be written as $H=H_0+H_1$, where
\begin{align}
    H_0 = \hbar\bar{\omega}_r a^{\dagger}a + \hbar g(a^{\dagger} + a)S_x + \hbar DS_x^2,
\end{align}
and $ H_1 = \bar{\omega}_qS_z$. We see that $H_0$ commutes with $S_x$ and it can be diagonalized by a \emph{polaron} transformation, $H_0^{\prime} =U^\dag H_0 U$, where $     U = \exp\left( -\gamma(a^{\dagger} - a)S_x \right)$ and $\gamma = g/\bar \omega_r $.
We obtain
\begin{equation}
H_0^{\prime} =   \hbar \bar \omega_r a^\dag a  + \hbar \underbrace{\left(D - \frac{g^2}{\bar \omega_r}\right)}_{\delta}   S_x^2.
\end{equation}
Therefore, in this new frame the eigenstates are $|s,m_x\rangle|n\rangle$, where $s$ is the total spin quantum number, $m_x$ is the spin projection along $x$ and $|n\rangle$ is the number state of the resonator mode. The corresponding energies are $E_{m_x,n}=\hbar  \bar\omega_r n+ \hbar\delta m_x^2$. After transforming back into the original frame we obtain the eigenstates $|\Psi_{s,m_x,n}\rangle$ defined in Eq.~\eqref{eq:Eigenstates}.

For $\delta\rightarrow 0$ the eigenspectrum of $H_0$ consists of a set of highly degenerate manifolds separated by multiples of $\hbar \bar \omega_r$. This degeneracy is lifted by the qubit Hamiltonian $H_1$ and in the following we are interested in the effect of $H_1$ on the ground-state manifold spanned by the states with $n=0$. To do so we need the matrix elements 
\begin{align}\label{eq:HqElements}
  &\langle \Psi_{s,m_x^\prime,n}| H_1 |\Psi_{s,m_x,0}\rangle =\\ \nonumber
  &\quad\hbar \bar \omega_q   e^{-\frac{\gamma^2(m_x-m_x^\prime)^2}{2}}   \frac{Ê\gamma^n (m_x^\prime-m_x)^n }{\sqrt{n!}} \langle s,m_x^\prime | S_z   |s,m_x\rangle.
\end{align}
To first order in $H_1$ we 
obtain a tunneling between neighboring $m_x$ states within the $n=0$ manifold, i.e.,  
\begin{equation}
 \langle \Psi_{s,m_x^\prime,0}| H_1   |\Psi_{s,m_x,0}\rangle =  \hbar \bar \omega_q   e^{-\frac{\gamma^2}{2}} \langle s,m_x^\prime | S_zÊ   |s,m_x\rangle.
\end{equation} 
In terms of the effective spin operators $\tilde S_k$ 
we can write the first order correction to the effective ground state Hamiltonian as
\begin{equation}
H_{\rm eff}^{(1)}= \hbar \bar \omega_q   e^{-\frac{\gamma^2}{2}}   \tilde S_z. 
\end{equation} 

To second order in $\bar \omega_q$ the states in the ground state manifold are coupled to higher $n$-states, which are separated by an energy $\hbar \bar \omega_r n$. These processes can be treated in second order perturbation theory and we obtain
\begin{equation}
H_{\rm eff}^{(2)} =   \sum_{s = 0}^{N/2} \sum_{m_x,m_x^\prime = -s }^{s}  M(s,m_x^\prime,m_x) Ê |\Psi_{s,m_x^\prime,0}\rangle\langle \Psi_{s,m_x,0}|,
\end{equation} 
where 
\begin{equation}
\begin{split}
&M(s,m_x^\prime,m_x)= \\
& - \sum_{n = 1}^{\infty}  \frac{\langle \Psi_{s,m_x^\prime,0}|H_1|\Psi_{s,m_x+1,n}\rangle  \langle \Psi_{s,m_x+1,n}|H_1|\Psi_{s,m_x,0}\rangle}{(2m_x+1) \hbar\delta + \hbar\bar \omega_r n}\\
&-\sum_{n = 1}^{\infty}  \frac{\langle \Psi_{s,m_x^\prime,0}|H_1|\Psi_{s,m_x-1,n}\rangle  \langle \Psi_{s,m_x-1,n}|H_1|\Psi_{s,m_x,0}\rangle}{ (-2m_x+1) \hbar\delta + \hbar \bar \omega_r n}.
\end{split}
\end{equation}
As a main contribution 
we obtain a diagonal term, which for small $\delta$ and $\gamma \gtrsim 2$ is approximately given by
\begin{equation}
\begin{split}
M(s,m_x,m_x) 
= &\frac{\hbar \bar \omega_q^2}{2\bar \omega_r}  \left[ m_x^2 - s(s+1)\right] e^{-\gamma^2} \sum_{n = 1}^{\infty}  \frac{\gamma^{2n}}{n!n} \\
\simeq & \frac{\hbar \bar \omega_q^2}{2\bar \omega_r\gamma^2} \left[ m_x^2 - s(s+1)\right],
\end{split}
\end{equation}
and we obtain the second order contribution 
\begin{equation}\label{eq:Heff2}
H_{\rm eff}^{(2)} = \frac{\hbar \bar \omega_q^2 \bar \omega_r}{2 g^2}  \left[ \tilde S_x^2 - \vec{\tilde S}^2 \right].
\end{equation}
Since these energy shifts are not exponentially suppressed, $H_{\rm eff}^{(2)}$ dominates over $ H_{\rm eff}^{(1)} $ 
and determines the basic ordering of the energy levels for $g\rightarrow \infty$. In particular, this result shows that for given $m_x$ the maximal angular momentum state with $s=N/2 $ is lowest in energy. Since $H$ preserves the total angular momentum, it is enough to evaluate the low excitation properties within this $s=N/2 $ manifold. 

Note that in addition to the energy correction we also obtain a correction to the state vectors. For most quantities these corrections are not essential, but they can lead to additional contributions in the expectation values, that are not taken into account in the analysis in the main text. A significant correction occurs, for example, for the qubit polarization $\langle \Psi_{N/2,0,0}| S_z |\Psi_{N/2,0,0}\rangle$, which according to $H_{\rm eff}$ would decay exponentially at large $g$, but in reality decays algebraically, i.e., $\langle \Psi_{N/2,0,0}| S_z |\Psi_{N/2,0,0}\rangle  \approx - N(N+2) \bar \omega_q\bar \omega_r/(4g^2)$.

Finally, we emphasize that the validity of the effective Hamiltonian $H_{\rm eff}$ requires that for each $n$ the matrix elements given in Eq.~\eqref{eq:HqElements} are small compared to the energy difference $\hbar \bar \omega_r n$. The matrix elements are exponentially suppressed for small $n$ and reach maximal value for $n_0\approx \gamma^2$. Therefore, the validity of the perturbation theory is restricted to parameters
\begin{equation}
\frac{\sqrt{s(s+1)}\bar \omega_q }{2} < \frac{g^2}{\bar \omega_r }.
\end{equation}


\end{document}